# Single-Electron Capacitance Spectroscopy of Individual Dopants in Silicon

M. Gasseller[†], R. Loo[‡], J. F. Harrison[§], M. Caymax[‡],
S. Rogge[∥] and S. H. Tessmer[†],*

[†] Department of Physics and Astronomy, Michigan State University, East Lansing, MI 48824, USA
[‡] IMEC, Kapeldreef 75, B-3001 Leuven, Belgium
[§] Department of Chemistry, Michigan State University, East Lansing, MI 48864, USA
[∥] Kavli Institute of Nanoscience, Delft University of Technology, Lorentzweg 1, 2628 CJ Delft, The Netherlands

* To whom correspondence should be addressed: tessmer@msu.edu

## ABSTRACT

Motivated by recent transport experiments and proposed atomic-scale semiconductor devices, we present measurements that extend the reach of scanned-probe methods to discern the properties of individual dopants tens of nanometers below the surface of a silicon sample. Using a capacitance-based approach, we have both spatially-resolved individual subsurface boron acceptors and detected spectroscopically single holes entering and leaving these minute systems of atoms. A resonance identified as the $B^+$ state is shown to shift in energy from acceptor to acceptor. We examine this behavior with respect to nearest-neighbor distances. By directly measuring the quantum levels and testing the effect of dopant-dopant interactions, this method represents a valuable tool for the development of future atomic-scale semiconductor devices.

As the size of conventional semiconductor components is reduced to nanometer scales, the exponential Moore's-Law improvement in the performance is determined by ever fewer numbers of dopants. The ultimate goal is to develop devices based on manipulating the charge and spin of individual dopant atoms[1,2]. Elucidating the electronic structure of these minute systems is a difficult technical challenge.

In a recent study, Kuljanishvili and coworkers performed single-electron capacitance measurements of subsurface silicon donors in gallium-arsenide using Subsurface Charge Accumulation (SCA) imaging[3,4]. The method essentially probed the quantum states of clusters of donors by measuring the electron addition energies, the energy at which the system will accommodate an additional charge. This study did not have sufficient spatial resolution to resolve the charging of individual subsurface dopants. In another recent study, Caro and coworkers applied resonant tunneling spectroscopy to study the quantum states of boron acceptors in silicon[5]. A clear conductance resonance was observed and shown to be consistent with tunneling through the $B^+$ state, for which two holes are bound to the boron atom. With regard to the energy of the resonance, Caro *et al.* suggested that the proximity of neighboring boron atoms on average resulted in a significant shift in the energy of the $B^+$ state. In this paper we extend the SCA method to study the addition energy of *individual* dopants inside a host silicon semiconductor, with the goal of observing the behavior of the $B^+$ state at the single-acceptor level.

Figure 1a shows a schematic of the SCA method and the layout of our experiment. The key component is a sharp metallic tip connected directly to a charge sensor that achieves a sensitivity of 0.01 $e/\sqrt{Hz}$ [6]. We monitor the tip's AC charge $q_{tip}$ in response to an AC excitation voltage $V_{exc}$ applied to an underlying electrode. If the quantum system below the tip can accommodate additional charge, the excitation voltage causes it to resonate between the system and the underlying electrode – giving rise to an enhanced capacitance, $C = q_{tip}/V_{exc}$ [7]. To acquire an image we record $C$ while scanning the tip laterally over a plane with the tip's apex a distance of ~1 nm above the silicon surface, and with a fixed DC bias voltage $V_{tip}$. For capacitance spectroscopy, the tip's position is fixed (i.e. not scanned); we then sweep $V_{tip}$ to acquire a $C$-$V$ curve. For the measurements shown here, we used a chemically-etched PtIr tip prepared to have a nanometer-scale radius of curvature of the apex; the data are consistent tip radius less than 25 nm. All data presented here were acquired at a temperature of $T$=4.2 K. Tip voltages are reported with respect to the nulling voltage; this accounts for the tip-sample contact potential[3].

The silicon sample we employed contained a narrow layer of dopants, i.e. a "δ layer", 15 nm below the exposed surface. The sample was grown by chemical vapor deposition on a Si(001) substrate – similar to the wafer used by Caro et al. The layers consist of (starting from the bottom) $p^{++}$ Si (500 nm) / $p$ Si (20 nm) / $\delta$ / $p$ Si (15 nm). Boron is the dopant for the layers and the $\delta$ spike. The $p^{++}$ layer is degenerately doped ($N_B$=10$^{19}$ cm$^{-3}$) and serves as the underlying metallic electrode. The lower $p$ layer ($N_B \sim 10^{17}$ cm$^{-3}$) serves as the tunneling barrier, whereas the upper $p$ layer is simply a spacer. The $\delta$ spike has areal density of $\rho$=1.7x10$^{11}$ cm$^{-2}$. Fig. 1b shows the doping profile in the structure measured with secondary ion mass spectroscopy. Each of the layers is clearly discernible, with the $\delta$ spike appearing as a peak about 2 nm wide and with a maximum concentration of $N_B$=5x10$^{17}$ cm$^{-3}$. Fig. 1c shows the profile of the valence band edge, including the vacuum gap and tip. The tunneling barrier height of ~10 meV is formed by the valence band contribution to the bandgap narrowing[8]. The curved arrow indicates a charge tunneling on and off an acceptor state. For consistency we will refer to the charges as holes. In reality the measurement is not sensitive to the sign of the charge carrier; moreover, conceptually a hole entering an acceptor is equivalent to an electron exiting the state.

A key parameter is the average spacing between nearest-neighbor acceptors. Statistically, nearest-neighbor distances for objects randomly positioned in two-dimensions follow Poissonian distributions, as shown in the Supporting Information, Figure S1[9]. For areal density $\rho=1.7 \times 10^{11}$ cm$^{-2}$, the mean distance to the first neighbor is expected to be 12.1 nm. This length is comparable to the 15 nm δ layer depth, which sets the lower limit to the lateral resolution of the SCA method[10]. Hence the goal of resolving the charging of individual subsurface acceptors is within reach.

Figure 2a shows a representative capacitance image. The triangles and the black dot mark 16 local maxima. The $1.7 \times 10^{11}$ cm$^{-2}$ boron density implies that we should on average observe 14 dopants in an area of this size. Several similar capacitance images were acquired at different locations and we consistently find that the number of bright features approximately agrees with the expected number of boron acceptors. Moreover, comparisons to topographical images of the surface, acquired by using our method in tunneling microscopy mode[6] show no correlation between the surface topography and the capacitance features. Hence we conclude that the bright features likely correspond to subsurface *B* dopants. Supporting-Information Fig. S2 examines the functional form of a bright capacitance feature.

Figure 2b shows a *C-V* curve acquired at the indicated location. The *C-V* curves consistently showed the two types of features indicated by the arrows. The green arrow marks a step structure that reproducibly occurred near $V_{tip}=-0.25$ V, with little dependence on the lateral position of the tip. The red arrow indicates a peak structure, which typically was found near $V_{tip}=-0.075$ V. These peaks shifted on the scale of tens of millivolts and often resembled multiple peaks, depending on the location of the probe.

The step structure likely results from the accumulation of charge in the surface potential well indicated in Fig. 1c (green arrow). Although a detailed description of the formation of this layer is complicated by the fringing nature of the tip's electric field, we can estimate the threshold potential using a parallel-plate picture. The first charge will enter the surface accumulation layer at a tip potential of $V_{tip}^{step} = (\phi_B + \Delta)/\alpha e$, where *e* is the elementary charge, $\Delta$ is the quantum level spacing between the bottom of the well and the first hole state to appear, and $\alpha$ is the voltage lever-arm that scales the applied voltage to account for the potential drop between the tip and the layer of interest within the sample. We expect $\alpha \sim 0.2$ directly below the apex[11], and $\phi_B=11.7$ meV [5]. With regard to $\Delta$, the accumulation layer will initially form a quantum dot below the apex of the tip of radius ~25 nm, determined mostly by the tip radius. From a simple estimate based on a quantum box of this size, and using a hole effective mass of half of the free electron mass, we find $\Delta \sim 10$ meV. These values give $V_{tip}^{step} \sim 100$ mV. As $V_{tip}$ increase further, the dot will grow in size and more charges will enter. As a result, a conducting surface layer is expected to form giving rise to the characteristic step-like *C-V* curve[3].

To analyze the peak structure (red arrow), we consider a quantum state of energy $\varepsilon$ of a dopant that is a distance *z* below the surface and laterally displaced from the tip's apex by *r* (see Fig. 1a). We again start from the parallel-plate expression for the resonance tip voltage for this state:

$$V_{tip}^{peak} = \frac{\phi_B - \varepsilon}{\alpha(z,r)e}. \qquad (1)$$

Here, in contrast to the expression for $V_{tip}^{step}$, we have generalized the lever arm to account approximately for the fringing pattern of electric field between the tip and sample. Essentially, a dopant a few nanometers below the surface and directly below the tip, $r = 0$, has a larger lever arm than a deeper and laterally displaced dopant. We account for this by allowing the parameter to be a function of $z$ and $r$. Of course $\alpha(z,r)$ also depends on the distance between the tip and surface, which is ~1 nm; this variation can be effectively absorbed into $z$. A detailed analysis of the electrostatics of the tip-sample system is given in Ref. 11.

Fig. 2c redisplays the *C-V* data with a linear background subtraction (gray line in Fig. 2b) and an expanded voltage scale. The solid curve shows a calculated fit based on single-electron capacitance peaks, which are essentially the convolution of semi-elliptical peaks with the derivative of the Fermi function[11]. As detailed in the Methods section, the fitting procedure superposes 4-6 single-electron peaks and invokes two free parameters for each peak, $\alpha(z,r)$ and $V_{tip}^{peak}$, which are varied to achieve the best fit. Essentially, the peak height must equal $e\alpha(z,r)$ due to the quantization of charge[11], allowing us to do a case-by-case estimate of the lever arm. We see that this procedure reproduces the measured curve well. We define the primary peak as the peak with the largest $\alpha(z,r)$, which we interpret as arising from charge entering an acceptor directly below the tip's apex; whereas the other peaks, which we designate as secondary peaks, correspond to the charging of more distant dopants with respect to the tip position.

To extract the values of $\varepsilon$ from our data, we apply Eq. (1) using the measured $V_{tip}^{peak}$ values and the $\alpha(z,r)$ parameters determined for each peak by the above fitting procedure. These measurements were reproducible over time scales of several hours and we estimate the uncertainty of the extracted $\varepsilon$ values to be ±1 meV, as shown in Supporting-Information Fig. S3. For the primary peak shown in Fig. 2c, the lever arm was determined to be $\alpha(z,0) = 0.156$ and the extracted energy from Eq. (1) was $\varepsilon = 0.2$ meV. By probing different areas on the sample, we have observed 20 peaks of sufficient quality to apply this procedure. Fig. 2d shows the resulting histogram of the extracted energies, which have a mean value of 4.3 ± 2.9 meV, where the uncertainty was determined by the standard deviation. This compares well to the average $B^+$ energies of 6.7 meV observed for a similar sample by Caro and coworkers. Hence considering both the *C-V* and capacitance image data, we conclude that we are observing the $B^+$ state of individual boron dopants; the peaks in the *C-V* curves correspond to the energies ($\varepsilon^+$) and the imaged high-capacitance bright spots mark the locations of the acceptors.

To examine the effects of interactions of neighboring acceptors, we consider the secondary peaks more carefully. As shown in Ref. 11, the lever arm for a subsurface dopant displaced from the tip's apex by $r$ is proportional to the Lorentzian $[1+(r/w)^2]^{-1}$, where $w$ is approximately equal to the depth of the underlying conducting layer, which is 35 nm for our sample. If we assume that the primary peak arises from a dopant directly below the apex, we can normalize the Lorentzian with respect to corresponding lever arm, $\alpha_P = \alpha(z,0)$. Moreover, if we assume that for a given curve the primary and secondary dopants lie in the same horizontal plane, we can neglect the $z$-dependence. Hence we can express approximately the $r$-dependence of the lever arm for the secondary dopants $\alpha_S(r)$ as

$$\frac{\alpha_S(r)}{\alpha_P} = \left[1 + \left(\frac{r}{35nm}\right)^2\right]^{-1}. \qquad (2)$$

Returning to Fig. 2c, consider the secondary peak marked by the gray arrow. For this peak, the best-fit lever arm is 0.097; thus $\alpha_S/\alpha_P = 0.097/0.156 = 0.622$. Applying Eq. (2), we assert that this dopant was approximating a distance of $r=27$ nm from the primary dopant. Supporting-Information Fig. S4 presents a table that shows the lever-arm parameters and the extracted distances of secondary dopants for all of the Fig. 2c peaks (and a similar analysis for data acquired at a nearby location). These distances compare well to the spacings of the bright features in the images; the comparison allows us to estimate the uncertainty for extracting distances in this way.

The resonant tunneling spectroscopy experiment by Caro *et al.* suggested that the proximity of neighboring acceptors resulted in a shift of the average $B^+$ binding energy to 6.7 meV, instead of the expected 2.0 meV [5]. Our method allows us to directly investigate the effect of neighboring dopants; we do this by examining the measured $\varepsilon^+$ of a particular primary peak versus the distance to its nearest neighbor, $R$, based on the analysis of secondary peaks.

We have obtained sufficiently high-quality data to extract $\varepsilon^+$ and $R$ values for six measurements. The results are shown in Fig. 3. We see that for three of the data sets the nearest-neighbor distance was less than 12 nm (the expected mean distance) and for two of the data sets $R$ was greater than 12 nm. The plot shows that acceptors with closer nearest neighbors tend to have greater binding energies for the second hole. This trend is consistent with the interpretation of the resonant tunneling spectroscopy experiment. Fig. 2d and Fig. 3 highlight the innovation of our scanning probe experiment: In contrast to most transport measurements which find the average behavior of many dopants, millions in the case of Ref. 5, we are able to discern both the average and the <u>atom-by-atom</u> binding energies.

To explore theoretically the effect of neighboring dopants on the $B^+$ state, we have calculated the energy to add a third hole to a system of two neutral acceptors separated by a distance $R$. The calculations were performed using the configuration-interaction method[12,13] in the context of the effective-mass theory[14,15]. In this approximation, an acceptor is regarded as a hydrogenic atom with an effective Rydberg energy $Ry^*$ and Bohr radius $a_0^*$. Both of the effective parameters are defined with respect to the first hole of an isolated acceptor: $Ry^*$ is equal to the binding energy and $a_0^*$ is equal to the average radius of the wave function. For boron in silicon $Ry^*=45.7$ meV, and we estimate $a_0^*=1.6$ nm, following Ref. 5. Fig. 3 (inset) shows the calculated third-hole binding energy in red. For self-consistency we retain to notation $\varepsilon^+$ for the energy of this state; however the hole is clearly interacting with both dopants, as discussed below. For $R>9a_0^*$, we show the asymptotic solution of $\varepsilon^+$, which equals that of an isolated acceptor with two holes[16]. We see that $\varepsilon^+$ is predicted to be positive for separations $R>3a_0$, indicating that the third hole is bound. In the range of separations $5a_0^*<R<16a_0^*$ we see that the calculated curve has a small negative slope. Our data are consistent with this trend, and qualitatively consistent with the resonant tunneling measurements. However the data show greater-than-predicted slope compared to our model. Moreover, the calculation shows a maximum near $R=5a_0^*$ which is not apparent in the measured data.

The behavior for large separations, $R>5a_0^*$, can be understood as a consequence of basic quantum-mechanics of molecules. Consider $R>>10a_0^*$, in which case each $B^0$ is effectively isolated; here it is well established that an additional hole experiences a weak attraction to each neutral acceptor due to polarization. Hence either one can bind the third hole and form the $B^+$ state. This state is analogous to hydrogen-minus, a proton with two bound electrons. For the isolated $B^+$ state the size of the wave function is rather large as the root-mean-square distance of the hole to the acceptor is $r^+ = 5.8a_0^*$ [5]. If we now allow the two neutral acceptors to be separated by $R \sim 10a_0^*$, the third hole has some reasonable probability to be found in either location. Hence, the corresponding wave function resembles a molecule with two peaks centered on each atom, each peak with a width $\sim r^+$. As the acceptor distance decreases the gradient of the wave function decreases along the line connecting the two atoms. The effect decreases the energy of the state, which increases the binding energy of the third hole. At sufficiently close separations this behavior will break down as the system will be poorly described as a two-acceptor molecule, instead behaving more as a single atom[17]. Our calculations show that this crossover should occur for $R \cong 5a_0^*$, where $\varepsilon^+$ exhibits the maximum.

The level of agreement between our measurements and calculated curve is not surprising in light of the highly simplified nature of the theoretical model, which incorporates only one neighbor and neglects effects arising from the semiconductor host such as the periodic crystalline potential and the multivalley structure of the Si conduction band. Indeed, we expect the effects of the host crystal to become more pronounced for dopants separated by less than a few nanometers, in which case $R$ becomes comparable to several lattice constants. Such considerations may remove the non-monotonic behavior predicted by our hydrogenic model. To the best of our knowledge, at present there is no complete theory for the binding energy of the third hole (or electron) as a function of separation for a semiconductor dopant pair. However, recent calculations by the Das Sarma group focusing on one electron bound to closely spaced pairs of P donors in Si predict that the precise separation and orientation of the dopant pair with respect to the crystalline axes are crucial parameters in determining the ground state wave functions[18,19].

**Acknowledgment.** We gratefully acknowledge insights provided by C. Piermarocchi and T. A. Kaplan, and helpful conversations with M. I. Dykman, B. Golding, C. Kayis, I. Kuljanishvili, S. D. Mahanti and M. Y. Simmons. This work was supported by the Michigan State Institute for Quantum Sciences, the Nation Science Foundation, DMR-0305461, DMR-0906939 and the Dutch Organization for Fundamental Research on Matter (FOM).

**FIGURES**

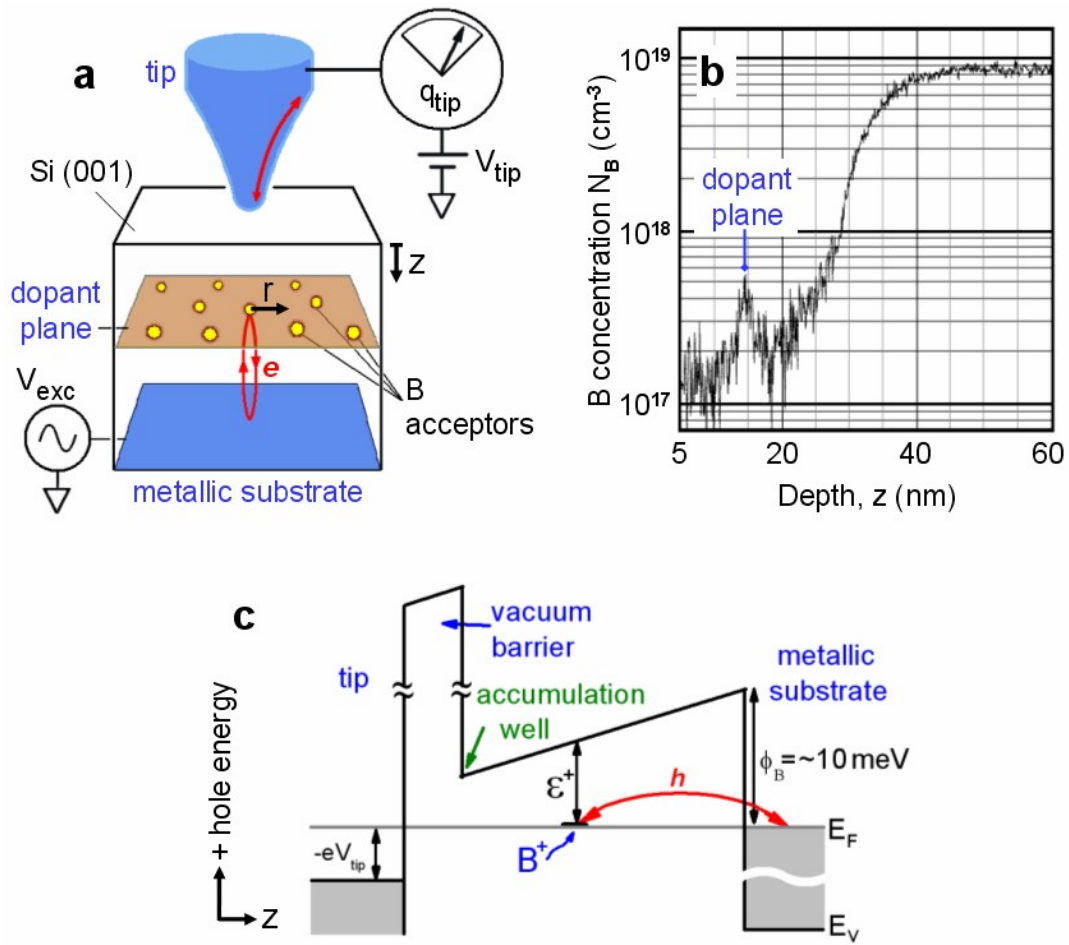

**Figure 1.** Capacitance-based scanning probe method to detect the charging of boron acceptors in a silicon sample. (a) Schematic of the Subsurface Charge Accumulation method and sample geometry. (b) Secondary ion mass spectroscopy (SIMS) measurement of the boron doping profile of the silicon sample. The top surface corresponds to $z=0$; SIMS data from $z=0-5$ nm exhibit a surface artifact and are not shown. (c) Inverted valence band profile, including the vacuum gap and tip. The red arrow indicates a hole tunneling on and off the $B+$ state of a boron acceptor.

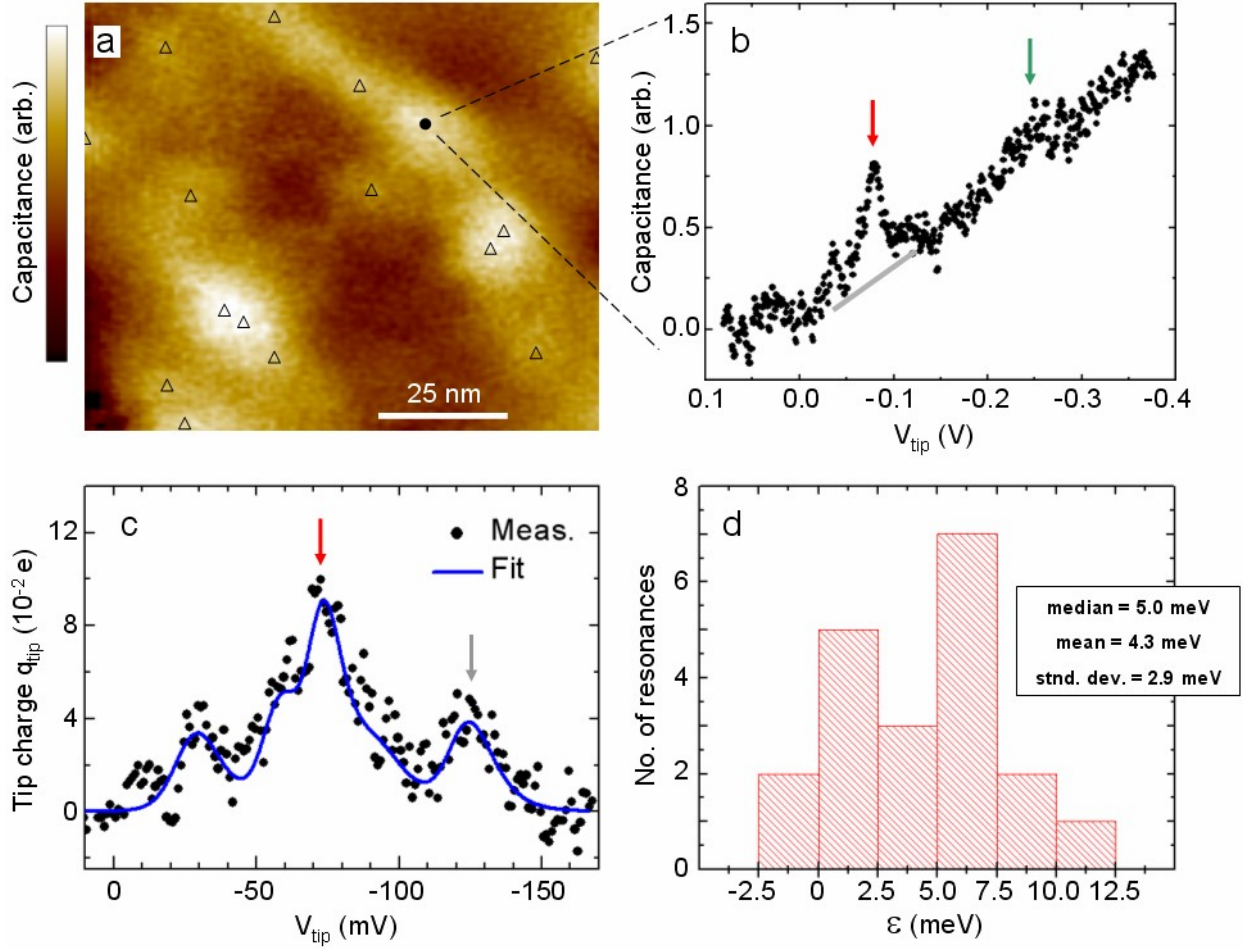

**Figure 2.** Representative capacitance data. (a) A capacitance image, acquired by fixing the voltage at $V_{tip}$= -0.075 V while scanning. Bright high-capacitance features are clearly resolved. (b) *C-V* curve acquired at the indicated location of image (a). Two types of features were consistently observed here and throughout the sample: peaks near $V_{tip}$=-0.075 V (red arrow) and a step near -0.250 V (green arrow). The peaks are consistent with single charges entering the dopant layer; the step is consistent with the formation of an accumulation layer at the sample surface. Additional curves and images are shown in the supporting figures. (c) Expanded *C-V* curve showing the peak structure from (b) with a background line (gray) subtracted. The blue solid curve is a fit from on a model based on the assumption that each peak corresponds to a single hole entering the system. In this case the red arrow indicates the primary peak, which we interpret as corresponding to a hole entering a dopant directly below the tip's apex. Smaller peaks are interpreted as holes entering more distant acceptors. (d) A histogram of the quantum state energies ε, extracted from measurements of 20 distinct peaks.

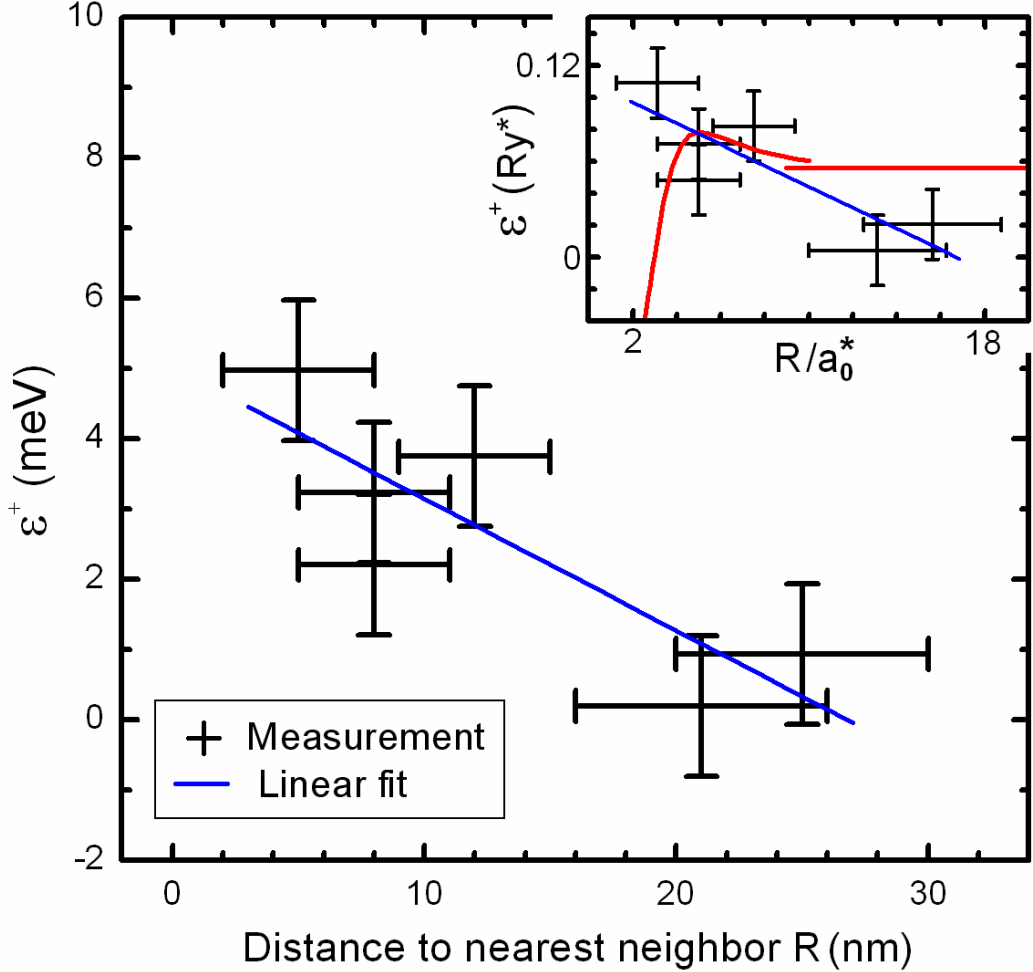

**Figure 3.** Hole binding energy versus nearest-neighbor distance. The main plot shows primary-peak binding energy plotted as a function of nearest-neighbor distance $R$. The parameters were extracted from measurements of dopants at different locations, including the data show in Fig. 2 (c). Shown in blue is a least-squares linear fit. The error bars show the estimated uncertainty of the energy and distance; details of these estimates are given in Supporting Information, Figure S3 and S4. The inset shows the data replotted with respect to the effective Rydberg energy and Bohr radius and compared to a hydrogenic model (red). The model incorporates two neutral acceptors and finds the binding energy of a third-hole; $\varepsilon^+$ is positive if the third charge results in a lower electronic energy for the system, i.e., if the third hole is bound. For $R/a_0^*>9$ in place of the calculation, we show the asymptotic solution of $\varepsilon^+$, which equals the solution for an isolated acceptor[16].

**Supporting Information**

**METHODS**

   **Data acquisition.** We begin each data run by scanning the tip in both tunneling and capacitance modes to check that the surface is sufficiently clean and free of major electronic defects[6]. To acquire the capacitance curves, we position the tip about 1 nm from the silicon surface and hold it at the fixed location while sweeping the tip voltage. The peaks seen in the charging signal show little lagging-phase structure and can be considered as essentially a change in capacitance.
   Our sensor circuit includes a bridge that allows us to subtract away the background mutual capacitance of the tip and sample, ~20 fF. Hence our plotted signal represents the change in capacitance as a function of voltage. All voltages are plotted with respect to the effective zero potential. This is the voltage for which no electric field terminates on the tip; it is shifted from the applied voltage by an amount equal to the contact potential, $V_{contact}$. $V_{contact}$ was measured directly with the same tip and sample using a modified Kelvin probe method[4] and found to be 225 ±10 mV. In the experiment presented here, the 10 mV uncertainty is scaled down by a factor of ten by the lever arm parameter; as a result, it is less significant than other sources of uncertainty.
   With regard to noise in the images and *C-V* curves, while the charge sensor itself attains a sensitivity of 0.01 $e/\sqrt{Hz}$, we find the effect of vibrations reduces our achieved noise level to about 0.1 $e/\sqrt{Hz}$. An experimental challenge for these measurements is the long data acquisition time required to achieve sub-single-hole sensitivity. Essentially the rate of data acquisition must be of order 1 s$^{-1}$. For the data presented here, we used a rate of 0.33 points per second for the *C-V* curves, and ~5 pixels per second for the capacitance images. In contrast to STM methods, we do not use feedback to maintain the tip position. This puts a stringent demand on the stability of the tip position. In Supporting Information Figure S3 we demonstrate the reproducibility of a measurement after several hours.
   **Fitting procedure.** With regard to the procedure used to generate the solid curves shown in Fig. 2, S3 and S4, it is instructive to briefly discuss the functional form of a single-electron peak. As discussed in detail in Ref. 11, for unfiltered data, four parameters determine the peak shape: the excitation amplitude $V_{exc}$, the temperature *T*, the local lever arm $\alpha$, and the peak voltage $V_{tip}^{peak}$. The measurement here utilized an amplifier output low-pass filter with time constant $\tau$, which tends to further broaden the peaks. This can be incorporated into a calculation by performing a simple integration. To generate the fitting curves, we used a modeling routine that superposes 4-6 single-electron peaks and then integrates the resulting curve appropriately. The excitation amplitude, temperature and time constant were fixed parameters, set to the experimental values of $V_{exc}$= 3.7 mV and *T*=4.2 K, $\tau$ =3.0 s. The parameters $\alpha$ and $V_{tip}^{peak}$ were considered as free and hence varied to achieve the best fit.

# NOTES ON SUPPORTING FIGURES

Here we remark on Supporting Figures S2 and S4; these remarks further explain the figures, in addition to the figure captions.

Figure S2 examines the functional form of an individual imaged dopant peak. As a first guess to the functional form of the peak, we compare the measured curve to a Lorentzian peak of width 15 nm: $L(r/15)=[1+(r/15 \text{ nm})^2]^{-1}$ (green curve). This is a logical guess since the spatial resolution of capacitance images of a subsurface conducting layer of depth $z$ is well-modeled invoking a convolution with the Lorentzian function, $L(r/z)$ [10,11]; in our case the average depth of the dopants is $z=15$ nm. However, clearly, the green curve gives a poor match to the measurement. We could extend this approach, and perform a fit to the Lorentzian shape by reducing the width $z$. But as SIMS measurements confirm the depth of the dopants to be 15 nm, it would be difficult to justify a characteristic peak function with a considerably smaller width parameter.

Surprisingly, simply squaring our first guess, $L^2(r/15)=[1+(r/15)^2]^{-2}$ yields a curve that provides an excellent fit, as shown in blue. This sharper-than-expected peak can be understood roughly using the following argument. Assume that at $r=0$, the tip is directly above the dopant and that $V_{tip}$ is set at the resonance voltage, as described by Eq. (1). Now as the tip is displaced laterally by $r$, two effects reduce the tip charge: Firstly, the response on the tip for a given charge entering the dopant is determined by their mutual capacitance, which is well-described by a Lorentzian function[11]; as $r$ increases, this effect causes $q_{tip}$ to decrease, following the Lorentzian form. Secondly, the potential at the dopant is scaled by the voltage lever arm, which also falls off with $r$ as a Lorentzian. Hence the resonance voltage is shifted away from $V_{tip}$, resulting in a further reduction of $q_{tip}$.

We have applied numerical simulations based on the boundary element method described in Ref. 11 to analyze in detail the expected form of an imaged single dopant. We find that the peak shapes are generally sharper than $L(r/z)$, and vary depending on a number of factors, including temperature and the excitation amplitude. These simulations show that it is somewhat fortuitous that the Lorentzian-squared function provides such a good fit in this case. Our modeling of the images will be the topic of a future publication.

Figure S4 examines curve fitting parameters and extracted distances of secondary dopants for data acquired at two locations. It is instructive to compare the distances of the triangles seen in the image, Fig. S4a, to the center of the respective circles and the distances extracted from the C-V curves shown in the bottom row of the table, Fig. S4c. Such a comparison shows that these distances are roughly consistent. The agreement is especially good for location B, for which the extracted distances are all 25 nm or less. In light of this approximate agreement, for nearest-neighbor distances $R$ shown in Fig. 3, we estimate the uncertainty is about ±3 nm for $R<15$ nm, and about ±5 nm for $R>15$ nm.

With regard to the interaction radius, following Ref. 11, we define the interaction radius $R_{int}$ with respect to the charging function, $Q(r)=L(r/R_{int})$; $Q(r)$ gives the image charge induced on the tip in response to a localized charge entering the dopant plane at position $r$. $R_{int}$ is expected to be approximately the average of the surface-to-dopant-plane distance and the surface-to-metallic-substrate distance, which is 25 nm for our sample.

**SUPPORTING FIGURES**

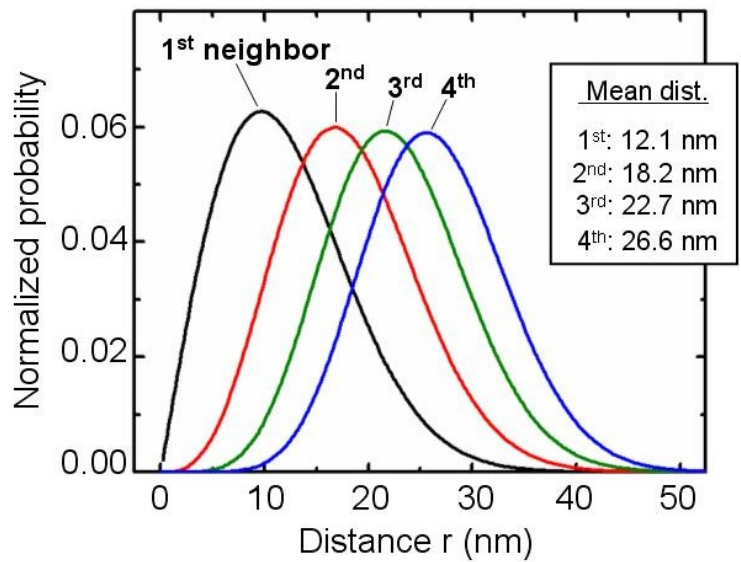

**Figure S1.** Statistical neighbor distances. Poissonian distributions of distances for a system of planar randomly positioned particles of average density $\rho=1.7\times10^{11}$ cm$^{-2}$. For a particle selected at random, the curves show the probability of finding the $m^{th}$ nearest neighbor at a distance of $r$, where $m=1,...,4$. The inset shows the mean nearest-neighbor distances. With regard to the 1$^{st}$ nearest neighbor, although $1/\sqrt{\rho}$ implies a distance of 24.3 nm, we see on average the distance to first neighbor is half this value at 12.1 nm.

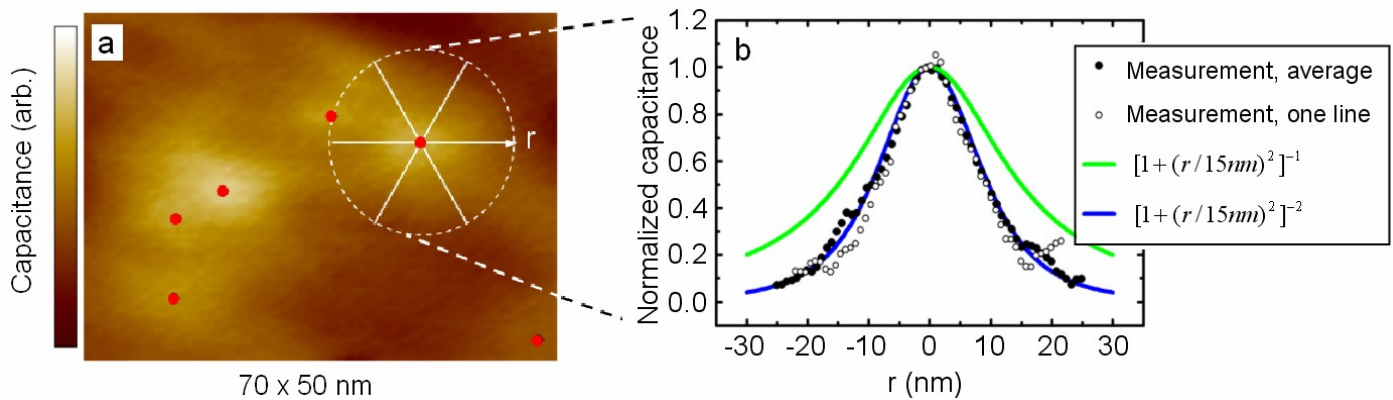

**Figure S2.** Functional form of an individual imaged dopant peak. (a) Capacitance image acquired with $V_{tip}$= -0.075 V. The data are unfiltered and minimally processed. Bright high-capacitance features are clearly present, as indicated by the red dots. Here focus on the bright spot near the upper right corner, which is relatively well isolated from its neighbors. (b) The average profile of the spot along three paths, as indicated. The average curve and the data along one of the individual paths are shown. As a first guess to the functional form of the peak, we compare the measured curve to a Lorentzian peak of width 15 nm: $L(r/15)=[1+(r/15 \text{ nm})^2]^{-1}$ (green curve). The fit is improved considerably by squaring the function, $L^2(r/15)= [1+(r/15)^2]^{-2}$ (blue curve). This sharper-than-expected peak can be understood as arising from two effects, as discussed in the notes, above.

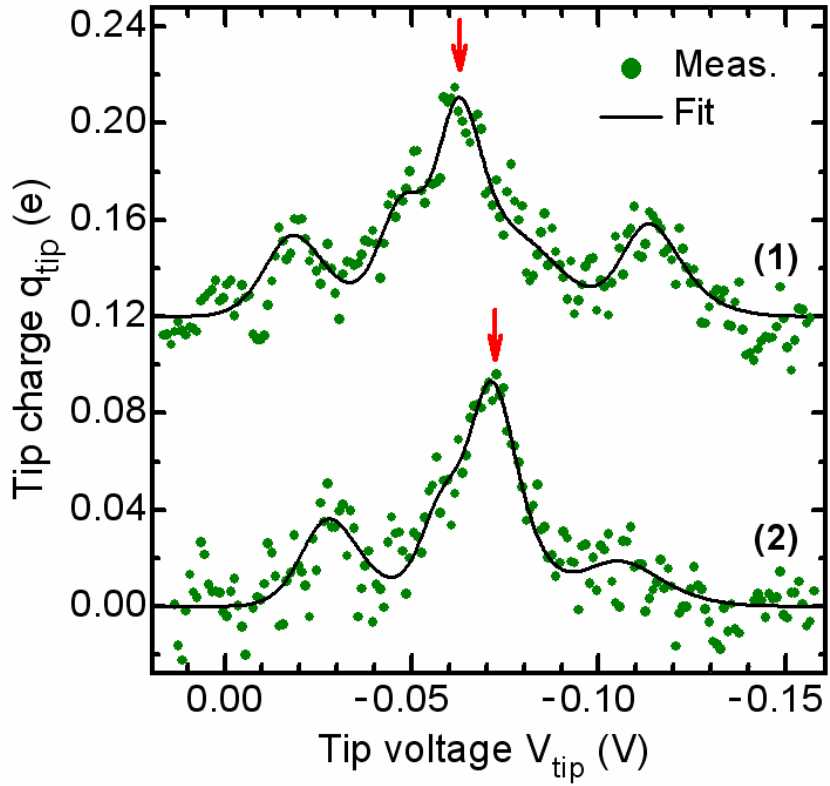

**Figure S3.** Reproducibility of C-V curves and estimate of energy uncertainty. As a test for reproducibility, here we show two curves labeled (1) and (2) are that were nominally acquired at the same location *A*, but with a time delay of 16 hours. (Curve (1) is the same data shown in Fig. 2(c).) We see that the curves have similar features, but the amplitudes and voltages of the peaks have changes somewhat. We attribute these changes to drift of the actual tip location, which alters the voltage lever arm for each dopant. Applying Eq. (1) to each curve to extract the energy of the acceptor state corresponding to the primary peak, we find $\varepsilon_1=0.2$ meV and $\varepsilon_2=-1.1$ meV; this comparison indicates that our analysis method is accurate to about $\pm 1$ meV.

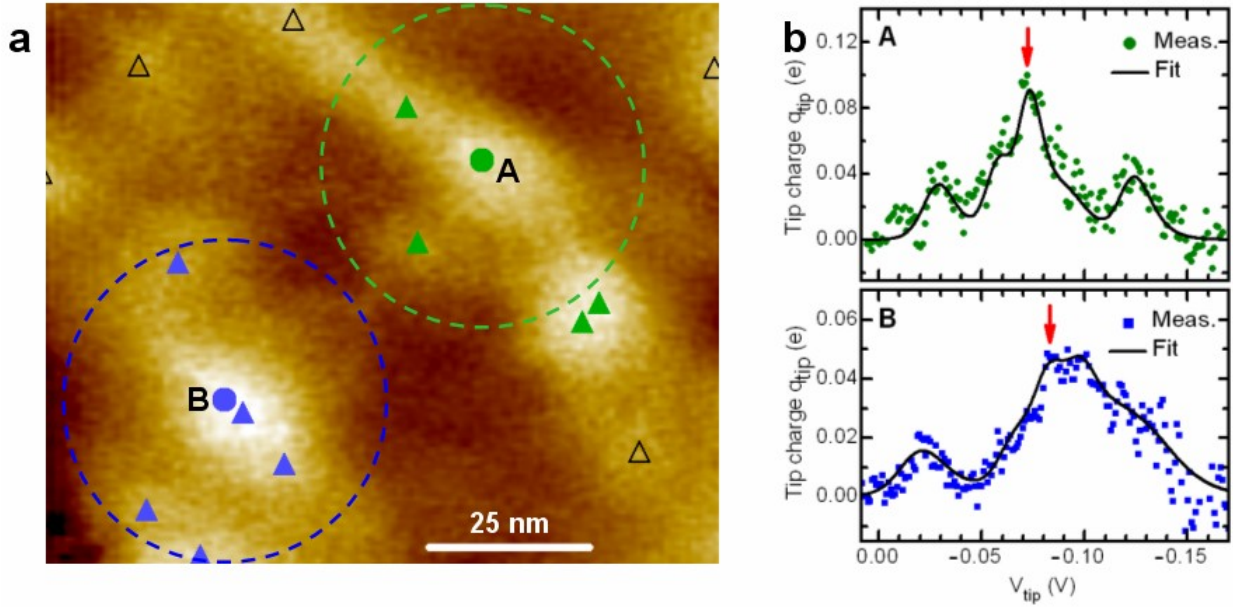

**Figure S4.** Comparison of curve fitting parameters, extracted distances of secondary dopants and estimate of nearest-neighbor-distance uncertainty. (a) Capacitance image reproduced from Fig. 2, with two indicated locations, *A* and *B*, where *C-V* curves were acquired. (*A* is the same location as that of Fig. 2 (b) and (c).) Triangles mark the likely locations of dopants; the hollow triangles correspond to dopants most likely too far from locations *A* and *B* to contribute to the respective *C-V* curves. The dashed circles are of radius 25 nm to show the approximate area of interaction for the tip at each location (see notes, above). (b) Capacitance spectra and fits for the two locations. To examine the details of the peaks, we have subtracted away a linear background from each curve, similar to the gray line shown in Fig. 2b. The vertical axis is converted to charge and displayed in units of e. (c) A table of the fitting parameters, $\alpha$ and $V_{tip}^{peak}$, for the *C-V* curves at locations *A* and *B*. The primary peak, *P*, corresponds to the resonance with the greatest lever arm; other peaks are referred to as secondary, *S*. The bottom two rows show the ratio of the fitted lever arms $\alpha_S/\alpha_P$ for each secondary peak and the estimated distances to the respective primary acceptor atom, based on Eq. (2).

17